\documentclass[twocolumn,secnumarabic,amssymb, nobibnotes, aps, prl]{revtex4-2}

\usepackage{makecell}
\usepackage{tabularx}
\usepackage{graphicx}
\usepackage{color}
\usepackage{dcolumn}
\usepackage{amsfonts}
\usepackage{amsmath}
\usepackage{bm} 
\usepackage{epstopdf}
\begin{document}
	
	
	\title{Experimental realization of a dusty plasma rocking ratchet with current reversal}
	
	
	
\author{Shun-xin Zhang$^{1}$}
\author{Shuo Wang$^{1}$}
\author{Ting-yu Yao$^{1}$}
\author{Yong-liang Zhang$^{1}$}
\author{Bao-quan Ai$^{2}$}
\email[Email:]{aibq@scnu.edu.cn}	
\author{Ya-feng He$^{1}$}
\email{heyf@hbu.edu.cn}
\affiliation{$^1$Hebei Research Center of the Basic Discipline for Computational Physics, College of Physics Science and Technology, Hebei University, Baoding 071002, China\\
	$^2$Key Laboratory of Atomic and Subatomic Structure and Quantum Control (Ministry of Education), Guangdong Basic Research Center of Excellence for Structure and Fundamental Interactions of Matter, School of Physics, South China Normal University, Guangzhou 510006, China.	  }
	
	
	\date{\today}
	
	\begin{abstract}
             A single dust particle confined in an asymmetric ratchet potential is periodically driven by two oppositely directed laser beams, forming an underdamped dusty plasma rocking ratchet. We experimentally investigate the transport dynamics of the particle under varying driving amplitudes and frequencies. Depending on the driving conditions, the particle exhibits positive, zero, or negative net currents, and current reversal is observed when the driving parameters cross critical thresholds. To interpret these transport behaviors, we develop a simplified model based on the competition between the driving force and the ratchet confinement. The model reveals that directional transport is governed by two requirements: the driving force must exceed the depinning threshold, and the duration of a driving semicycle must be longer than the uphill escape time from a ratchet well. The resulting dynamic phase diagram quantitatively reproduces the experimentally observed transport regimes and current reversals. These results demonstrate dusty plasma as a versatile platform for investigating nonequilibrium transport phenomena of underdamped particles in rocking ratchets.
	\end{abstract}
	
	\pacs{ 52.27.Lw }
	
	
	\maketitle
	
\subsection{I. INTRODUCTION}
\indent Ratchet systems generate directed transport under nonequilibrium conditions by combining spatial asymmetry with unbiased fluctuations or periodic driving forces. Depending on the driving mechanism, ratchets can be classified into several categories, including rocking \cite{Roca,HEPRE,Reichhardt}, flashing \cite{Kim,LiH,Roeling}, and tilted \cite{Kurths,Mateos} ratchets. Among them, rocking ratchets are particularly attractive because a time-periodic force with zero mean can induce a finite particle current by breaking spatiotemporal symmetry \cite{Gupta}. The resulting transport strongly depends on the driving amplitude, frequency, and the potential asymmetry. Rich dynamical phenomena, including directional transport \cite{Sven Matthias}, current reversal \cite{LiW}, synchronization \cite{FanLM}, and chaos \cite{ZhangPJ}, have been predicted and observed in various systems, such as optical lattices \cite{Magda}, semiconductor nanostructures \cite{Grossert}, cold atoms \cite{Kenfack}, and nanofluidic devices \cite{Skaug,Camacho}.

\indent Dusty plasmas provide a unique platform for studying ratchet transport because the motion of individual particles can be directly visualized and manipulated \cite{Morfill,Shukla,POPREV,Chu,Wang,Thomas,Ott,Killer,Du,Feng,Bajaj,Joshi,HuangD}. Micron-sized particles immersed in plasma typically acquire a large negative charge, allowing their motion to be manipulated by externally controlled electric or optical forces. Moreover, asymmetric ratchet potentials can be readily generated using sawtooth-shaped electrode structures \cite{He}, and the potential barrier height can be tuned through the discharge conditions. In addition, laser radiation pressure offers a convenient means of applying periodic driving forces with controllable amplitudes and frequencies. These unique advantages enable the realization of a highly tunable rocking-ratchet system for underdamped particles. Here, we experimentally realize such a system by alternately illuminating a dust particle with two oppositely directed laser beams. In the absence of an external force, a single dust particle remains trapped in the ratchet potential. When a periodic external force with zero average is applied by alternating the illumination of the two beams, the dust particle undergoes directional transport along the ratchet potential. Importantly, current reversal of the dust particle is achieved by changing the driving frequency and amplitude of the external force, as we investigate experimentally in this work.

\subsection{II. DUSTY PLASMA ROCKING RATCHET}
  \begin{figure}[htp]
	\centering
	\includegraphics[width=85mm]{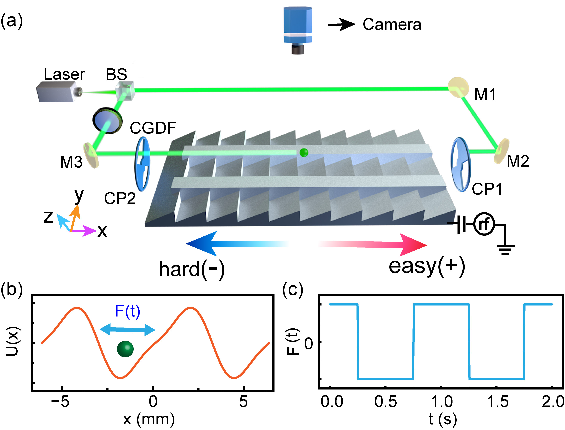}
	\caption{
		(color online). Experimental realization of a dusty plasma rocking ratchet. (a) Schematic of the experimental setup. A dust particle is confined in a sawtooth channel above a grooved lower electrode and driven by two counter-propagating laser beams. (b) The particle experiences an asymmetric ratchet potential $U(x)$ with easy and hard transport directions. (c) A zero-mean square-wave driving force $F(t)$ is generated using two phase-complementary choppers to alternate laser irradiation, with a circular gradient density filter (CGDF) adjusted to equalize the beam intensities. Directional transport and current reversal of the dust particle in the ratchet potential are realized by adjusting the driving force. (BS) beam splitter, (M1-M3) mirrors, (CP1-CP2) choppers, and (rf) radio-frequency power.}
	\label{figure1}
\end{figure}

\indent The experimental setup is schematically illustrated in Fig.~1(a). The powered lower electrode consists of a stainless steel plate featuring sawtooth-shaped grooves, electrically connected to a radio-frequency (13.56 MHz) power supply through a matching network. These grooves have identical geometric parameters: a depth of $h$ = 3 mm and a period length of $L$ = 5 mm. Argon gas is introduced into the chamber (not shown) at a flow rate of 2 sccm to minimize neutral gas perturbation. Argon plasma is then generated through capacitively-coupled discharge. Above the grooves, a non-electroneutral sheath forms with a spatial profile that smoothly follows the sawtooth-shaped grooves. As a consequence of the geometric asymmetry of the grooves, the potential distribution within the non-electroneutral sheath along the $x$- direction exhibits ratchet potential features \cite{Wang1,supple}, as exemplified by the red solid curve in Fig.~1(b). 

\indent As a single polystyrene microsphere (dust particle) with a radius of 13 $\mu$m is introduced into the plasma, it is trapped by the ratchet potential of the sawtooth-shaped grooves of the lower electrode. Two additional parallel straight metal strips are placed along the $x$- direction above the lower electrode. This spatial configuration creates a well-defined sawtooth channel as well as an asymmetric and periodic ratchet potential in the $x$- direction, thereby enabling systematic investigation of the rocking ratchet effect of a dust particle confined to one-dimensional motion along the sawtooth channel under a periodic driving force.

\begin{figure*}[htp]
	\begin{center}\includegraphics[width=17.5cm,height=6.95cm]{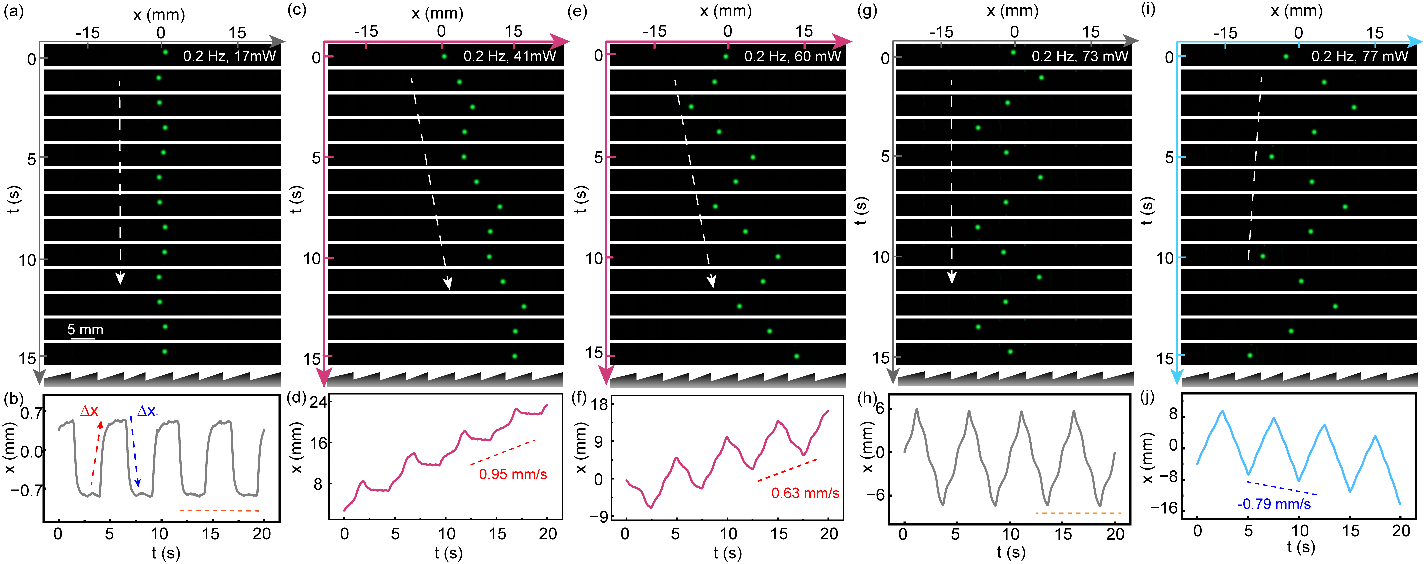}
		\caption{(color online). Directional transport and current reversal of a single dust particle in the dusty plasma rocking ratchet with increasing laser power at a fixed driving frequency of $f=0.2$ Hz. (a,b) Zero current ($P_L=17$ mW); (c,d) positive current ($P_L=41$ mW); (e,f) positive current ($P_L=60$ mW); (g,h) zero net current ($P_L=73$ mW); and (i,j) negative current ($P_L=77$ mW). Panels (a), (c), (e), (g), and (i) show representative snapshots of the dust particle moving along the sawtooth channel, while panels (b), (d), (f), (h), and (j) show the corresponding particle displacement as a function of time. The transport currents obtained from linear fits are 0, 0.95, 0.63, 0, and $-0.79$ mm/s, respectively. In panel (b), $\Delta x_+$ and $\Delta x_-$ denote the distances traveled by the dust particle along the easy and hard directions, respectively, during successive driving semicycles. The brightness and size of dust particles in the images are enhanced so that they can be seen more clearly.}
	\end{center}
\end{figure*}

\indent The periodic driving force is applied by irradiating the suspended dust particle with a pair of superimposed oppositely directed laser beams ($\lambda = 532$ nm) of equal intensity along the sawtooth channel. The strength $F$ of the laser force is controlled by adjusting the laser power, and the frequency $f$ is modulated by the rotation rate of a pair of phase-complementary choppers. The time-dependent laser forces have zero average:
\begin{eqnarray}
	F(t) = \left\{ \begin{array}{ll}
		F, & \textrm{$t$ $\in$ (0, T/2]}\\
		-F, & \textrm{$t$ $\in$ (T/2, T]}
	\end{array},  \right. 
\end{eqnarray}
where $T$ is the period of the laser force. The formula for $F$ is provided in Appendix A. For a trapped dust particle, the static depinning thresholds of the ratchet potential—defined as the minimum force required to propel a stationary dust particle out of the ratchet potential well—are measured to be $F_{R} = 0.51$ pN in the easy direction (right) and $F_{L} = 1.15$ pN in the hard direction (left). Here, $F_{L} > F_{R}$ due to the asymmetry of the ratchet potential. Using laser-excited resonances \cite{Konopka}, the measured fundamental frequency of the dust particle at the well bottom of the ratchet potential is $f_0 = 26.9$ Hz and the damping rate is $\gamma = 28.6$ s$^{-1}$, both of which are much larger than the driving frequency ($f < 2$ Hz in the experiments).

\indent The discharge conditions are fixed at a gas pressure of $p = 32$ Pa and a discharge power of $P = 20$ W throughout the paper. Experimental processes are monitored in real time with a top-view camera (EOS R5, $4096\times2160$ pixels, 30 fps).

\subsection{III. TRANSPORT AND CURRENT REVERSAL OF DUST PARTICLE} 
\indent As the main result of this paper, we realize directional transport and current reversal of a dust particle in our dusty plasma rocking ratchet experiment. The observed phenomena are strongly dependent on the amplitude and frequency of the driving force, which we present one by one in the following. 

\indent Figure 2 first demonstrates the directional transport and current reversal of a single dust particle under varying laser power at a fixed driving frequency of $f$ = 0.2 Hz. The particle motion movies under these conditions are provided in the Supplementary Video. When the laser force remains below both depinning thresholds ($F$ $<$ $F_{R}$, $F_{L}$), exemplified by $F$ = 0.36 pN at laser power $P_{L}$ = 17 mW [Figs.~2(a) and 2(b)], the dust particle is completely confined within the ratchet potential well, exhibiting oscillation with amplitude smaller than the sawtooth length, $\Delta x_{\pm}$ $<$ $L$. Therefore, no macroscopic directional transport is observed. 

\begin{figure}[htbp]
	\begin{center}\includegraphics[width=8cm,height=7.4cm]{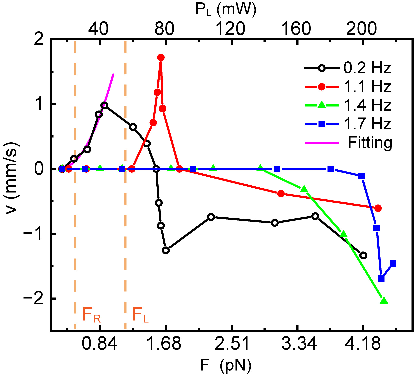}
		\caption{(color online). Current of particle transport as a function of the driving force at different laser powers. At lower driving frequencies ($f=0.2$ and $1.1$ Hz), current reversal from positive to negative is observed. At higher driving frequencies ($f=1.4$ and $1.7$ Hz), only zero and negative currents appear. $F_R$ and $F_L$ indicate the depinning forces.}
	\end{center}
\end{figure}
\indent As the laser power increases to intermediate values  ($F_{R}$ $<$ $F$ $<$ $F_{L}$, e.g. $F$ = 0.86 pN at $P_{L}$ = 41 mW), a novel transport regime emerges: the dust particle undergoes unidirectional jumps toward the easy direction during $F(t)$ $>$ 0 phases while remaining pinned in the hard direction [$F(t)$ $<$0], as evidenced by the asymmetric displacement trajectories quantified by $\Delta x_+$ $=$ 7.30 mm (forward) and $\Delta x_-$ $=$ 2.24 mm (backward) [Figs.~2(c) and 2(d)]. This asymmetric dynamics results in a positive net current of particle transport along the sawtooth channel. Linear fitting of the $x-t$ curve in Fig.~2(d) yields a steady-state transport velocity of $v$ = 0.95 mm/s. Notably, the dust particle exhibits metastable states on the steep side of the ratchet potential, due to $F$ $<$ $F_{L}$, manifested as plateaus in the $x-t$ curve of Fig.~2(d).

\indent When the laser force is further increased to exceed both the depinning thresholds ($F$ $>$ $F_{R}$, $F_{L}$), as exemplified by $F$ = 1.25 pN at $P_{L}$ = 60 mW [Figs.~2(e) and 2(f)], the dust particle becomes capable of depinning in both directions. Here, a remarkable feature of this regime is that the traveling distances of the dust particle in both the hard and easy directions exceed the length of a single sawtooth ($\Delta x_{\pm}$ $>$ $L$) within a half driving cycle. In this case, the additional depinning in the hard direction counteracts the forward transport, resulting in a reduction in the positive net current ($v$ = 0.63 mm/s) in the easy direction.

\indent At a specific laser force $F$ = 1.53 pN corresponding to $P_{L}$ = 73 mW, the traveling distances of the dust particle in the two directions within a half driving cycle become nearly identical, $\Delta x_+$ $=$ $\Delta x_-$ $=$ 13.43 mm (both exceeding $2L$), suggesting zero net current of particle transport, as demonstrated in Figs.~2(g) and 2(h).

\indent Upon further increasing the laser force (e.g. $F$ = 1.61 pN at $P_{L}$ = 77 mW), the traveling distances of the dust particle toward the hard direction exceeds that toward the easy direction, $\Delta x_-$ $>$ $\Delta x_+$, and $\Delta x_{\pm}$ $>$ $3L$, as shown in Figs.~2(i) and 2(j). This results in a negative net current of particle transport ($v$ = ${-}$0.79 mm/s) toward the hard direction.  

\indent Figure 2 collectively reveals that at the fixed driving frequency $f = 0.2$ Hz, the macroscopic velocity evolves through three distinct regimes with increasing laser force: it first increases from zero to a maximum positive value, then decreases to zero, and finally reverses to a negative value, as indicated by the black curve in Fig.~3. In particular, in this low-frequency (adiabatic) limit, the macroscopic velocity exhibits a quadratic scaling with laser force at low forces, $v \sim F^2$, as indicated by the solid black and pink fitting curves in Fig.~3, confirming the theoretical prediction for adiabatic ratchet transport [see Appendix B, Eq. (24)].

\indent The directional transport of the dust particle in the dusty plasma rocking ratchet is also strongly dependent on the driving frequency, as illustrated by the four frequency cases presented in Fig.~3. For a driving frequency of $f = 1.1$ Hz, the red $v$–$F$ curve exhibits a profile similar to the low-frequency ($f = 0.2$ Hz) case, with current reversal also observable. A key distinction is that the parameter window of laser force $F$ required for positive transport shifts to higher values; i.e., achieving positive current demands a larger $F$ for $f = 1.1$ Hz than for $f = 0.2$ Hz. At higher frequencies ($f = 1.4$, $1.7$ Hz), the transport behavior changes dramatically: no positive current is observed, and the current transitions directly from zero to negative as $F$ increases. Notably, the critical laser force required for this transition is larger at $f = 1.7$ Hz than at $f = 1.4$ Hz, indicating a frequency-dependent threshold for the onset of negative current.

\subsection{IV. MODEL} 
\indent The dynamics of a dust particle in the rocking ratchet are inherently complex, arising from the interplay of inertia, asymmetric potential modulation, and periodic driving. During transport, four distinct dynamical processes occur: uphill and downhill motion along both the easy and hard directions, each characterized by its own timescale. In addition, the ratchet-induced force and neutral drag vary throughout the driving cycle. To capture the essential transport behavior, we introduce a minimal model that focuses on the dominant mechanisms.
\begin{figure}[htp]
	\begin{center}\includegraphics[width=8cm,height=3.75cm]{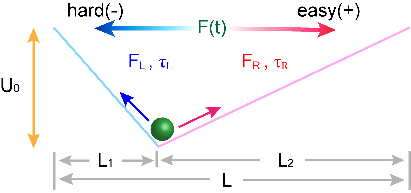}
				\caption{(color online). Simplified piecewise ratchet potential. $U_0$ and $L$ denote the barrier height and spatial period, respectively, and $L_1<L_2$ indicates the asymmetry of the ratchet potential. $F_L$ and $F_R$ are the constant depinning forces along the hard and easy directions, respectively, while $\tau_L$ and $\tau_R$ represent the corresponding uphill times. The green ball illustrates the dust particle.}		
	\end{center}
\end{figure}

\indent The key factor controlling transport is the particle's ability to escape the ratchet potential under an external driving force, while nonessential effects are neglected. The simplified model is based on three assumptions: (1) the asymmetric ratchet potential is represented as a piecewise linear function (Fig.~4),

\begin{eqnarray}
	U(x) = \left\{ \begin{array}{ll}
		U{_0}x/L_1 , & \textrm{$x$ $\in$ (0, $L_1$]}\\   
		U_0(L-x)/L_2 ,  & \textrm{$x$ $\in$ ($L_1$, $L$]}
	\end{array},  \right. 
\end{eqnarray}
where $U_0$ and $L$ denote the barrier height and spatial period of the ratchet potential, respectively, with $L=L_1+L_2$. The condition $L_1<L_2$ characterizes the spatial asymmetry of the potential and gives rise to two constant depinning forces, $F_L$ and $F_R$, corresponding to the hard and easy directions, respectively, with $F_L>F_R$. (2) The friction force is neglected. (3) We consider only the escape dynamics from a single ratchet well. Since directional transport is primarily determined by barrier crossing, the uphill motion is retained whereas the downhill motion is neglected. The particle is assumed to start from the potential minimum with zero initial velocity.

\indent Under these assumptions, the uphill time $\tau_L$ of the dust particle required to escape from the potential well along the hard direction under a left-directed driving force $F(t)$ = $-F$ is

\begin{eqnarray}
	\tau_L &=& \sqrt{\frac{2mL_1}{F-F_L}},
\end{eqnarray}
here, $m$ is the mass of dust particle. Similarly, the uphill time $\tau_R$ of the dust particle along the easy direction under the right-directed driving force $F(t)$ = $F$ reads,
\begin{eqnarray}
	\tau_R &=& \sqrt{\frac{2mL_2}{F-F_R}}.
\end{eqnarray}
Escape requires two conditions: (i) the driving force exceeds the depinning force, $F > F_{L,R}$, and (ii) the driving semicycle is longer than the uphill time, $T/2 > \tau_{L,R}$. When both conditions are met, the net current direction is determined by the shorter uphill time: $\tau_R < \tau_L$ gives a positive current, whereas $\tau_R > \tau_L$ results in a negative current.

\begin{figure}[htp]
	\begin{center}\includegraphics[width=8cm,height=11.9 cm]{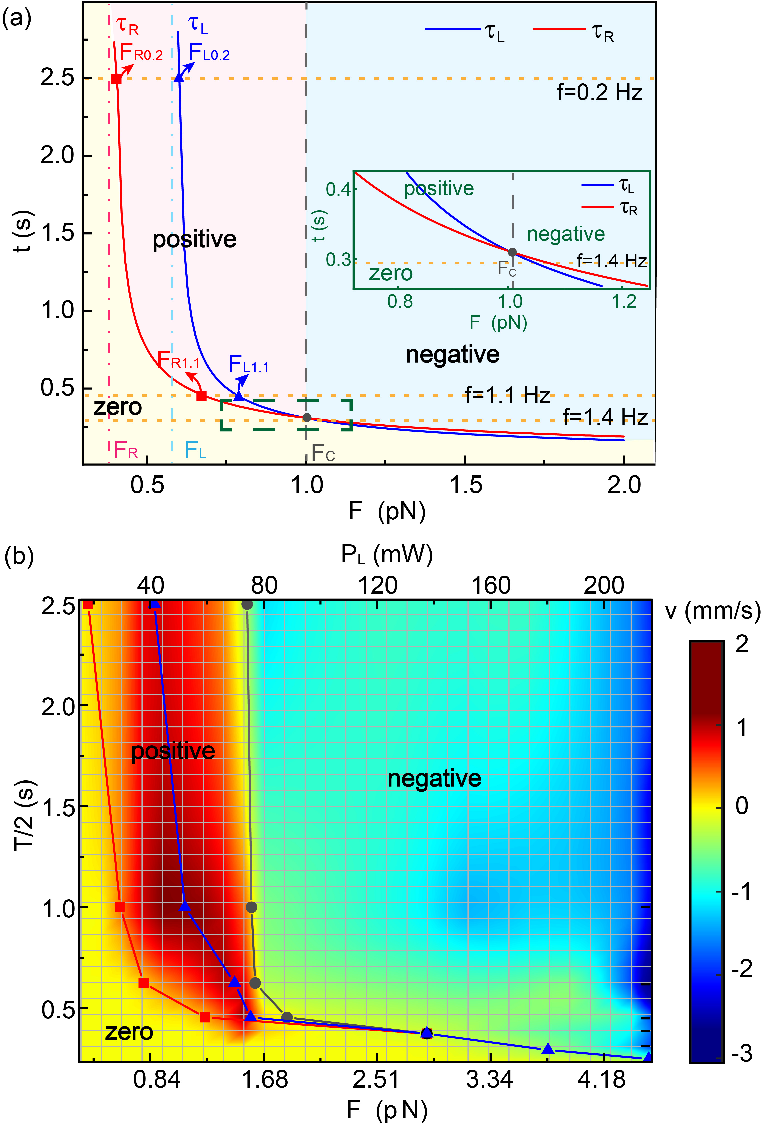}
		\caption{(color online). Transport phase diagram in the parameter space of driving force $F$ and time $t$. (a) Theoretical prediction based on the simplified model. The red and blue curves denote the uphill times $\tau_R$ and $\tau_L$ for depinning along the easy and hard directions of the ratchet potential, respectively. Their intersection defines the critical driving force $F_c$ (inset). Three transport regimes are identified: zero current, positive current, and negative current. The horizontal dashed lines indicate the experimental conditions corresponding to $f=0.2$, 1.1, and 1.4 Hz. (b) Experimental phase diagram, where the vertical axis represents the driving semicycle $T/2$. The red (square) and black (circle) curves separate the phase diagram into three regimes (zero, positive, and negative current), and the blue (triangle) curve marks the driving parameters that lead to the maximum positive current.}
	\end{center}
\end{figure}

\indent From Eqs. (3) and (4), both uphill times, $\tau_R$ and $\tau_L$, decrease monotonically with increasing driving force $F$. Using the parameters obtained from our simulations \cite{supple}, $U_0=-0.085$ V, $L_1=2$ mm, $L_2=3$ mm, and $m=9.66\times 10^{-12}$ kg, we construct the theoretical phase diagram in the $(F,t)$ parameter space, as shown in Fig.~5(a). The two curves, $\tau_R(F)$ and $\tau_L(F)$, intersect at a critical driving force $F_c$. For $F<F_c$, the uphill time along the easy direction is shorter than that along the hard direction, i.e., $\tau_R<\tau_L$. In contrast, for $F>F_c$, the relation is reversed and $\tau_R>\tau_L$.

\indent The phase diagram can therefore be divided into three distinct transport regimes. (i) A zero-current region (yellow), where either $F<F_R$ or the driving semicycle is too short for barrier crossing, namely $T/2<\min(\tau_R,\tau_L)$. (ii) A positive-current region (red), where $F_R<F<F_c$ and $T/2>\tau_R$, such that the particle preferentially escapes along the easy direction. (iii) A negative-current region (blue), where $F>F_c$ and $T/2>\tau_L$, leading to preferential transport along the hard direction. The theoretical phase boundaries are in good agreement with those obtained experimentally [Fig.~5(b)]. The resulting dynamic phase diagram captures the essential competition between the driving force and the escape dynamics of the particle, thereby providing a simple physical picture for the experimentally observed transport transitions and current reversals.

\indent For a driving frequency of $f=0.2$ Hz, corresponding to a driving semicycle of $T/2=2.5$ s, the transport behavior can be understood by following the upper dashed line in Fig.~5(a). When $F<F_{R0.2}$, the particle remains trapped in the ratchet well and no net current is generated. As the driving force increases to $F_{R0.2}<F<F_{L0.2}$, the condition $T/2>\tau_R$ is satisfied, allowing the particle to depin only along the easy direction. Consequently, a positive current emerges and increases with increasing $F$. For $F_{L0.2}<F<F_c$, the particle becomes capable of depinning in both directions. The additional motion along the hard direction partially compensates the forward transport, resulting in a reduction of the positive current. When the driving force exceeds the critical value $F_c$, the escape time along the hard direction becomes shorter than that along the easy direction ($\tau_L<\tau_R$). The particle therefore depins preferentially along the hard direction, giving rise to a negative current and hence a current reversal. Overall, the evolution from zero to positive and subsequently negative current reflects the competition between the driving semicycle and the directional escape times, reproducing the experimentally observed current reversal at $f=0.2$ Hz.

\indent For a driving frequency of $f$ = 1.1 Hz, the system exhibits the same qualitative sequence of transport regimes as in the low-frequency case ($f$ = 0.2 Hz). As $F$ increases, the current changes from zero to positive and finally becomes negative. Nevertheless, the positive-current region is noticeably compressed and shifted toward higher driving forces. Since the driving semicycle is shorter at higher frequency, a larger force is required to reduce the uphill time $\tau_R$ below $T/2$, thereby enabling depinning along the easy direction. Consequently, the interval of positive transport, $[F_{R1.1},F_c]$, becomes narrower than that for $f$ = 0.2 Hz. These theoretical predictions are in good agreement with the experimental results presented in Fig.~3.

\indent At higher driving frequencies, the transport dynamics undergo a qualitative change. For instance, at $f$ = 1.4 Hz (lowest dashed line in Fig.~5(a)), the driving semicycle becomes too short for the dust particle to escape the ratchet potential at moderate forces. Specifically, when $F < 1.15$ pN (see inset), even though the driving force exceeds the depinning thresholds $F_{R,L}$ and the critical force $F_c$, the condition $T/2 > \tau_L$ is not satisfied, preventing depinning. As the driving force increases further and $T/2$ surpasses $\tau_L$, the particle begins to depin along the hard direction ($\tau_L < \tau_R$), resulting in a negative net current. This mechanism, which produces a direct transition from zero to negative current, accurately accounts for the $v$–$F$ curves observed experimentally at $f$ = 1.4 and 1.7 Hz in Fig.~3.
	
\subsection{V. DISCUSSION AND CONCLUSION}
\indent The dusty plasma rocking ratchet exhibits robust operation over a broad range of experimental parameters beyond those used in the present study. Additional tests show that dust particles with diameter ranging from 5 to 26 $\mu$m undergo directional transport under gas pressures of 10–50 Pa and discharge powers of 20–50 W. Since particles of different sizes respond differently to the same driving force, the resulting transport behavior suggests potential applications in size-selective particle separation.

\indent The observed transport is largely independent of the dust particle’s initial position and velocity. We tested this by perturbing the particles prior to applying the driving laser force, finding consistent transport outcomes. While numerical simulations \cite{Zarlenga} and optical/SQUID rocking ratchet experiments \cite{Arzola,Arzola2,Zapata} have revealed Devil’s staircase structures and chaotic dynamics, such nonlinear phenomena may also emerge in dusty plasma ratchets under more precise experimental conditions.  

\indent Although the real dynamics involve repeated interwell hopping, downhill relaxation, and neutral drag, the transport direction is primarily determined by the first barrier-crossing event in each driving semicycle. Consequently, current generation is governed by two coupled requirements: the driving force must overcome the corresponding depinning force, and the driving semicycle must be sufficiently long for the particle to complete the uphill escape process. For analytical clarity, neutral drag is neglected in the first approximation. Since the experimentally measured damping time ($\gamma^{-1} \sim 0.035$ s) is much shorter than the driving period, drag mainly affects the quantitative values of the uphill times without qualitatively altering the phase boundaries.
	
\indent In summary, we experimentally realized a dusty plasma rocking ratchet by periodically driving a single dust particle confined in a sawtooth-shaped channel with alternating laser radiation pressure. Depending on the driving amplitude and frequency, the particle exhibits positive, zero, and negative transport currents, including current reversal. A simplified escape-time model reveals that the transport behavior is governed by the interplay between the driving force and the depinning threshold, together with the competition between the driving semicycle and the directional uphill times. The resulting phase diagram quantitatively reproduces the experimentally observed transport regimes. These results establish dusty plasma as a controllable platform for investigating nonequilibrium transport of underdamped particles and may facilitate future applications in particle manipulation and separation.
	
\subsection{ACKNOWLEDGMENTS}
\indent This work was supported by the National Key R\&D Program of China under Grant No. 2025YFF0512000, the National Natural Science Foundation of China (Grants No. 12275064, No. 12475203 and No. 12475036), the Natural Science Foundation of Hebei Province of China (Grant No. A2024201020). and the Project supported by the Space Application System of China Manned Space Program. 

\subsection{APPENDIX A: THE FORMULA FOR LASER RADIATION PRESSURE}

\indent This section focuses on the calculation method for the radiative pressure $F$ exerted on particles. The laser force \cite{LiuB}
\begin{eqnarray}
	F&=&q n_1 \pi r_{d}^{2} I_{L}/c,
\end{eqnarray}
where  $q$ is a dimensionless factor determining the absorption of photons, $n_1$ is the refractive index of the medium around the particle and $c$ is the speed of light.  $I_{L}$ represents the power density of the laser beam with $I_{L}$= $P_{L}$/($\pi$$r_{p}^2$), where $r_{p}$ = 0.18 mm is the radius of the aperture used for calibration. The laser force is proportional to the laser power $P_{L}$ and the cross-section of the dust particle $\pi r_{d}^{2}$.

\subsection{APPENDIX B: TRANSPORT MECHANISMS AT LOW-FREQUENCY AND HIGH-FREQUENCY LIMITS}
\indent Consider a particle of mass $m$ moving in a periodic potential $U(x)$ with period $L$, subjected to an external driving force $F(t)$ and thermal noise $\xi(t)$. The underdamped Langevin equation reads
\begin{eqnarray}
	\dot{x} &=& v ,
\end{eqnarray}
\begin{eqnarray}
	m\dot{v} &=& -m\gamma  v -U'(x) +  F(t)  + \sqrt{2m\gamma  k_B  T_a} {\xi(t)} ,
\end{eqnarray}
where, $\xi(t)$ is Gaussian white noise with $\langle$$\xi(t)$$\xi(t')$$\rangle$$=$$\delta(t-t')$, $\gamma$ the damping rate, $T_a$ the temperature, $k_B$ the Boltzmann constant.  

\indent For mathematical convenience, the square wave driving force in the experiment is theoretically approximated as a sinusoidal form $F$ = $A_0$sin($ft$), where $A_0$ and $f$ represent the driving force amplitude and angular frequency, respectively. Simultaneously, the periodic ratchet potential $U(x)$ 
\begin{eqnarray}
	U(x) &=& U_0[\sin(\frac{2\pi x}{L}) +\frac{1}{4}\sin(\frac{4\pi x}{L}) ] .
\end{eqnarray}

\indent  Let $P$($x$, $v$, $t$) be the probability density in phase space. The Kramers equation is
\begin{eqnarray}
	\begin{split}
	\frac{\partial P}{\partial t} + v \frac{\partial P}{\partial x} + \frac{1}{m} [-m\gamma  v -U'(x) \\
	+  A_0\sin (ft)  ]\frac{\partial P}{\partial v} = \frac{\gamma  k_B  T}{m}\frac{\partial^2 P}{\partial v^2} .
\end{split}
\end{eqnarray}

The marginal (position) probability density is
\begin{eqnarray}
	\rho (x,t) &=& \int _{-\infty}^{\infty} v P(x,v,t)  dv.
\end{eqnarray}
It satisfies the continuity equation
\begin{eqnarray}
	\frac{\partial  \rho (x,t)} {\partial t} + \frac{\partial  J_x (x,t)} {\partial x} &=& 0,
\end{eqnarray}
\begin{eqnarray}
	J_x (x,t) &=& \int _{-\infty}^{\infty} v P(x,v,t)  dv,
\end{eqnarray}
here $J_x (x,t)$ is the position probability current.

\indent  When barrier crossings are rare (Kramers regime), the net current may be expressed in terms of instantaneous transition rates $r_\pm (t)$ to the neighboring wells:
\begin{eqnarray}	
	v_{net}(t) & \simeq & L[r_+(t) - r_-(t)],
\end{eqnarray}
\begin{eqnarray}	
	\overline{J_x} & = & \frac{f}{2 \pi} \int _{0}^{2 \pi/f} v_{net}(t) dt.
\end{eqnarray}

\indent For weak tilting by $F$($t$), barrier heights are approximated as
\begin{eqnarray}	
	\Delta U_+(F) & \approx & \Delta U_0 - a_+F,
\end{eqnarray}
\begin{eqnarray}	
	\Delta U_-(F) & \approx & \Delta U_0 + a_-F,
\end{eqnarray}
where $a_\pm$ denote effective distances from the minimum to the right/left saddle points. Then
\begin{eqnarray}	
	r_+(t) & \approx & r_{0+} \exp[-(\Delta U_0 - a_+F(t))/k_B T_a],
\end{eqnarray}
\begin{eqnarray}	
	r_-(t) & \approx & r_{0-} \exp[-(\Delta U_0 + a_-F(t))/k_B T_a].
\end{eqnarray}

\indent  {\bf(A) Low-Frequency (Adiabatic) Limit} $f$ $\to$ 0. The system follows the instantaneous rates:
\begin{eqnarray}
	\begin{split}	
	\overline{J_x}^{(adiabatic)}  = & \frac{f}{2 \pi} \int _{0}^{2 \pi/f} L [r_+(A_0 \sin ft) 	 \\  
	-& r_-(A_0 \sin ft)] dt.
\end{split}	
\end{eqnarray}
Expanding to second order in $Ao$ yields (with  $\Delta a^2 $ $ \equiv $ $a^2_+$ $-$ $a^2_-$, $r^*$ $=$ $r_0$ $\exp[-\Delta U_0 /k_B T_a]$ :

\begin{eqnarray}	
	\overline{J_x}^{(low)} &  \approx  & L r^* \frac{\Delta a^2}{4(k_B T_a)^2} A^2_0.
\end{eqnarray}
The current scales as $A^2_0$ and is frequency-independent in the adiabatic limit.

\indent  {\bf (B) High-Frequency Limit }  $f$ $\gg$ $\gamma$. At large $f$, the particle cannot follow the rapid oscillations, the rectification effect is suppressed. A convenient interpolation form introduces a frequency-response factor:
\begin{equation}
	\begin{split}	
	\overline{J_x}^{(high)}   \approx  & L r^* \frac{\Delta a^2}{4(k_B T_a)^2} \frac{A^2_0}{1+(  f / \gamma )^2}  \\
	  \sim & L r^*\frac{\Delta a^2}{4(k_B T_a)^2} \frac{\gamma ^2 A^2_0}{f^2},  (f \gg \gamma). 
   \end{split}
\end{equation}
Thus at high frequency the net current decays as $1/f^2$ due to inertial suppression.


\begin{thebibliography}{}
		
		\bibitem{Roca} J. Martín-Roca, L. I. Solis, F. M. Pedrero, P. Casadejust, I. Pagonabarraga, and C. Calero,	Colloidal Model for Investigating Optimal Efficiency in Weakly Coupled Ratchet Motors, Phys. Rev. Lett. $\mathbf{135}$, 028301 (2025).
		\bibitem{HEPRE} Y. F. He and B. Q. Ai, Enhancement of the longitudinal transport by a weakly transversal drive, Phys. Rev. E $\mathbf{81}$, 021110 (2010). 
	    \bibitem{Reichhardt} C.	Reichhardt and C. J. O. Reichhardt, Stripe and bubble ratchets on asymmetric substrates, Phys. Rev. Res. $\mathbf{6}$, 043290 (2024).	
	
	   
	  
    	\bibitem{LiH} H. Li, T. Gao, and S. J. Xie, Flashing-ratchet effect on directional photocurrent in organic devices, Phys. Rev. Appl. $\mathbf{23}$, 024020 (2025).
	    \bibitem{Kim} D. K. Kim and H. Jeong, Deep reinforcement learning for feedback control in a collective flashing ratchet, Phys. Rev. Res. $\mathbf{3}$, L022002 (2021).	
	    \bibitem{Roeling} E. M. Roeling, W. C. Germs, B. Smalbrugge, E. J. Geluk, T. de Vries, R. A. J. Janssen, and M. Kemerink, Organic electronic ratchets doing work, Nat. Mater. $\mathbf{10}$, 51–55 (2011).  
	    \bibitem{Kurths} Y. G. Li, Y. Xu, and J. Kurths, Roughness-enhanced transport in a tilted ratchet driven by Levy noise, Phys. Rev. E $\mathbf{96}$, 052121 (2017).
	    \bibitem{Mateos} J. L. Mateos, and F. R. Alatriste, Phase synchronization in tilted inertial ratchets as chaotic rotators, Chaos  $\mathbf{18}$, 043125 (2008). 
	    
	   
	   

       \bibitem{Gupta} A. Gupta and P. S. Burada, Separation of interacting active particles in an asymmetric channel, Phys. Rev. E $\mathbf{108}$, 034605 (2023). 
       \bibitem{Sven Matthias} S. Matthias and F. Müller, Asymmetric pores in a silicon membrane acting as massively parallel Brownian ratchets, Nature (London), $\mathbf{424}$, 53-57 (2003). 
	  \bibitem{LiW} W. Li, D. Huang, C. Reichhardt, C. J. O. Reichhardt, and Y. Feng, Bidirectional flow of two-dimensional dusty plasma under asymmetric periodic substrates driven by unbiased external excitations, Phys. Rev. Res. $\mathbf{5}$,  023008 (2023).
	   \bibitem{FanLM} L. M. Fan, M. G. Li, T. F. Gao, and J. D. Bao, Multiple modulations of coupling effects on directed transport of Brownian particles driven by nonequilibrium fluctuations, Phys. Rev. E $\mathbf{112}$, 024135 (2025). 
	  \bibitem{ZhangPJ} P. J. Zhang, J. Q. Zhang, P. Wang, J. Huo, and X.M. Wang, Directed transport of two-coupled particles under the coordination of the coupling and an asymmetric potential, Chaos Solitons Fractals $\mathbf{182}$, 114830 (2024).
	  
	  
	  
	   \bibitem{Magda} M. G. Sánchez-Sánchez, R. de J. León-Montiel, and P. A. Quinto-Su, Phase Dependent Vectorial Current Control in Symmetric Noisy Optical Ratchets, Phys. Rev. Lett. $\mathbf{123}$, 170601 (2019).
	   \bibitem{Grossert}  C . Grossert, M. Leder, S. Denisov, P. Hänggi, and M. Weitz, Experimental control of transport resonances in a coherent quantum rocking ratchet,   Nat. Commun. $\mathbf{7}$, 10440 (2016)		
		\bibitem{Kenfack} A. Kenfack, On the skewness of optical lattice ratchet potentials transport of cold atoms at quantum resonance, Opt. Commun. $\mathbf{600}$, 132701 (2026).
		\bibitem{Skaug} M. J. Skaug, C. Schwemmer, S. Fringes, C. D. Rawlings, and A. W. Knoll, Nanofluidic rocking brownian motors, Science $\mathbf{359}$, 1505-1508 (2018).
		\bibitem{Camacho} G. Camacho, A. Rodriguez-Barroso, O. Martinez-Cano, J. R. Morillas, P. Tierno, and J. de Vicente, Experimental realization of a colloidal ratchet effect in a non-newtonian fluid, Phys. Rev. Appl. $\mathbf{19}$, L021001 (2023). 

		
		
		
		\bibitem{Morfill} G. E. Morfill and A. V. Ivlev, Complex plasmas: An interdisciplinary research field, Rev. Mod. Phys. $\mathbf{81}$, 1353-1404 (2009).
    	\bibitem{Shukla} P. K. Shukla and B. Eliasson, Colloquium: Fundamentals of dust-plasma interactions, Rev. Mod. Phys. $\mathbf{81}$, 25-44 (2009).
    	\bibitem{POPREV} J. Beckers, J. Berndt, D. Block, M. Bonitz, P. J. Bruggeman, L. Cou\"{e}del, G. L. Delzanno, Y. Feng, R. Gopalakrishnan, F. Greiner, P. Hartmann, M. Hor\'{a}nyi, H. Kersten, C. A. Knapek, U. Konopka, U. Kortshagen, E. G. Kostadinova, E. Kova\v{c}evi\'{c}, S. I. Krasheninnikov, I. Mann, et al., Physics and applications of dusty plasmas: The Perspectives, Phys. Plasmas $\mathbf{30}$, 120601 (2023).
	    \bibitem{Wang} Y. N. Wang, L. J. Hou, and X. G. Wang, Self-consistent nonlinear resonance and hysteresis of a charged microparticle in a rf sheath, Phys. Rev. Lett. $\mathbf{89}$, 155001 (2002).              
	    \bibitem{Chu} J. H. Chu and L. I, Direct observation of Coulomb crystals and liquids in strongly coupled rf dusty plasmas, Phys. Rev. Lett. $\mathbf{72}$, 4009-4012 (1994).
	    
	    
    	
    	\bibitem{Thomas} H. M. Thomas and G. E. Morfill, Melting dynamics of a plasma crystal, Nature (London) $\mathbf{379}$, 806-809 (1996).           
 		\bibitem{Ott} T. Ott, M. Bonitz, P. Hartmann, and Z. Donk\'{o}, Spontaneous generation of temperature anisotropy in a strongly coupled magnetized plasma, Phys. Rev. E $\mathbf{95}$, 013209 (2017).
    	\bibitem{Killer} C. Killer, T. Bockwoldt, S. Sch\"{u}tt, M. Himpel, A. Melzer, and A. Piel, Phase separation of binary charged particle systems with small size disparities using a dusty plasma, Phys. Rev. Lett. $\mathbf{116}$, 115002 (2016).
    	\bibitem{Du} C. R. Du, V. Nosenko, H. M. Thomas, Y. F. Lin, G. E. Morfill, and A. V. Ivlev, Slow Dynamics in a quasi-two-dimensional binary complex plasma, Phys. Rev. Lett. $\mathbf{123}$, 185002 (2019).
        \bibitem{Feng} D. Huang, M. Baggioli, S. Y. Lu, Z. Ma, and Y. Feng, Revealing the supercritical dynamics of dusty plasmas and their liquidlike to gaslike dynamical crossover, Phys. Rev. Res. $\mathbf{5}$, 013149 (2023).	
    
    		
    	
		\bibitem{Bajaj} P. Bajaj, S. Khrapak, V. Yaroshenko, and M. Schwabe, Spatial distribution of dust density wave properties in fluid complex plasmas, Phys. Rev. E $\mathbf{105}$, 025202 (2022).
		\bibitem{Joshi} E. Joshi, M. Y. Pustylnik, M. H. Thoma, H. M. Thomas, and M. Schwabe, Recrystallization in string-fluid complex plasmas, Phys. Rev. Res. $\mathbf{5}$, L012030 (2023).
		\bibitem{HuangD} D. Huang, S. Y. Lu, M. S. Murillo, and Y. Feng, Origin of viscosity at individual particle level in Yukawa liquids, Phys. Rev. Res. $\mathbf{4}$, 033064 (2022). 
		\bibitem{He} Y. F. He, B. Q. Ai, C. X. Dai, C. Song, R. Q. Wang, W. T. Sun, F. C. Liu, and Y. Feng,  Experimental demonstration of a dusty plasma ratchet rectification and its reversal, Phys. Rev. Lett. $\mathbf{124}$, 075001 (2020).
        \bibitem{Wang1} S. Wang, S. X. Zhang, X. Liu, T. Y. Yao, X. Z. Wang, B. Q. Ai, and Y. F. He, Separation of microspheres using a potential ratchet in dusty plasma, Appl. Phys. Lett. $\mathbf{127}$, 154103 (2025). 
		\bibitem{supple} See Supplemental Materials at XXXX for a detailed description of the numerical simulations and Supplementary Videos,  which includes Ref. [35]. 
		\bibitem{COMSOL} COMSOL Multiphysics, www.comsol.com.
		
		
		\bibitem{Konopka}  U. Konopka, G. E. Morfill, and L. Ratke, Measurement of the Interaction Potential of Microspheres in the Sheath of a rf Discharge, Phys. Rev. Lett. $\mathbf{84}$, 891-894 (2000).
		\bibitem{Zarlenga} D. G. Zarlenga, H. A. Larrondo, C. M. Arizmendi, and F. Family. Chaos in kicked ratchets, Phys. Rev. E $\mathbf{91}$, 032901 (2015).	
		 \bibitem{Arzola} A. V. Arzola, M. Villasante-Barahona, K. Volke-Sep\'{u}lveda, P. J\'{a}kl, and P. Zem\'{a}nek, Omnidirectional transport in fully reconfigurable two dimensional optical ratchets, Phys. Rev. Lett. $\mathbf{118}$, 138002 (2017).
		\bibitem{Arzola2} A. V. Arzola, K. Volke-Sepúlveda, and J. L. Mateos, Experimental Control of Transport and Current Reversals in a Deterministic Optical Rocking Ratchet, Phys. Rev. Lett. $\mathbf{106}$, 168104 (2011).
		\bibitem{Zapata} I. Zapata, R. Bartussek, F. Sols, and P. Hänggi, Voltage Rectification by a SQUID Ratchet, Phys. Rev. Lett. $\mathbf{77}$, 2292-2295 (1996). 	
		
		
	
		\bibitem{LiuB} B. Liu, J. Goree, V. Nosenko, and L. Boufendi, Radiation pressure and gas drag forces on a melamine-formaidehyde microsphere in a dusty plasma, Phys. Plasmas  $\mathbf{10}$, 9-20  (2003).	
		
		
		
		
		
		%
		
	\end{thebibliography}
\end{document}